\documentclass [notitlepage, floatfix, superscriptaddress, showpacs, aps,pra, twocolumn, 10pt]{revtex4-1}

\usepackage{amsmath}
\usepackage{hyperref}
\usepackage{graphicx}
\usepackage{bbm}
\usepackage{amssymb}
\usepackage[english]{babel}
\usepackage{lmodern}
\usepackage[T1]{fontenc}
\usepackage{wasysym}
\usepackage[normalem]{ulem}
\usepackage{float} 
\usepackage{dcolumn}
\usepackage{color}
\usepackage[caption=false]{subfig} 

\usepackage{amssymb}
\usepackage{gensymb} 

\graphicspath{{Figures/}}

\newcommand{\rosso}{}

\begin{document}

\title{A non-inductive magnetic eye-tracker: \\ from dipole tracking to gaze retrieval.}

\author{Valerio Biancalana} 
\email{valerio.biancalana@unisi.it}
\affiliation{DSFTA, University of Siena -- Via Roma 56, 53100 Siena, Italy}
\author{Piero Chessa}
\affiliation{Dept. of Physics "E.Fermi", University of Pisa, Largo Pontecorvo 3, 56127 Pisa, Italy}

\begin{abstract}
  We analyze the information that can be retrieved from the tracking parameters produced by an innovative wearable eye tracker. The latter is based on a permanent-magnet marked corneal lens and by an array of magnetoresistive detectors that measure the magnetostatic field in several positions in the eye proximity. We demonstrate that, despite missing information due to the axial symmetry of the measured field, physiological constraints or measurement conditions make possible to infer complete eye-pose data. Angular precision and accuracy achieved with the current prototypical device are also assessed and briefly discussed.  
The results show that the instrumentation considered is suitable as a new,  moderately invasive medical diagnostics for the characterization of ocular movements and associated disorders.
\end{abstract}

\nopagebreak
\maketitle

\section{Introduction}

The localization of magnetostatic sources on the basis of multiple measurements performed in known, pre-assigned positions is widely studied and
finds application in several areas, including medical diagnostics \cite{than_ieee_12, dinatali_ieee_13, juce_sens_22}. Setups based on permanent-magnet sources enable wireless measurements, which come with low invasivity and low cost. The developed methodology produces rich tracking data that include both target position and orientation and presents the advantage of an occlusion-free detection. Modern highly magnetized materials and solid-state magnetometric devices help to reduce the invasivity level and to improve the portability/wearability of the required instrumentation. We have recently demonstrated that this methodology can achieve sufficient time and space resolution to be applicable also to eye-tracking.

Concerning the instrumentation aimed to track ocular movements, devices with diverse degrees of invasivity, accuracy, speed, wearability have been developed on the basis of several concurrent technologies. The latter include electro-oculography\cite{kolder_oph_74}, infrared-reflection devices \cite{kumar_lar_92, aungsakun_ijps_12}, video cameras \cite{kimmel_fbn_12, houben_iovs_06}, and inductive magnetic receivers (scleral search coil, SSC) \cite{robinson_ieee_63, collewijn_vr_75, zee_on_14, zee_jov_15}. In this list, the former items are less precise and invasive while the latter is considered the top-performance at expenses of a noticeable invasivity. \rosso{It is worth noticing that the need of an electrical connection constitutes the main source of the SSC invasivity, and that  unwired (double induction) SSC have been proposed \cite{reulen_ieee_82, bour_ieee_84, bremen_jnm_07}, which lead to an invasivity level comparable to that of our non-inductive approach}.

Our non-inductive magnetometric instrumentation  for eye-tracking is characterized by an intermediate level of invasivity (much less than SSC), low intrusivity (it is fully wearable), low cost (<1 k\texteuro), high speed (>100 Sa/s),  high precision (< 1 degree) and good robustness with respect to external disturbances, such as eye-blinking, facial muscles actuation and fluctuations of ambient illumination. In addition, after a preliminary calibration of the sensors, it does not require additional (patient based) re-calibrations.

Apart from some works aimed to localize multiple  \cite{tarantino_ieee_20, gherardini_cmpb_21, ge_ieee_21} or distributed \cite{jaufenthaler_sens_20, tanwear_ieee_20} magnetic sources, most of the \rosso{non-inductive} magnetic trackers reported in the literature use some numerical algorithms to infer the magnetic target pose (position and orientation) modeling the magnet in terms of a dipolar field source. Our device does not make an exception to this common rule, but, differing from other setups, it tracks both the target pose and the ambient field. \rosso{To this end, both the magnet and the dipolar field of the magnet and the homogeneous field of the Earth are taken into account in the field modeling. Consequently the data analysis provides both the dipole pose and the ambient field,} making the latter no longer a disturbance term, but a source of additional information. 

Like other single-target systems, our device provides comprehensive dipole pose information, which however is not sufficient to fully characterize the target object's pose, even if simply modeled as a rigid body.
More explicitly, the information that can be extracted from dipole tracking includes three spatial and two angular co-ordinates (the system being blind to rotations around the dipole direction), thus only 5 out of the 6 degrees of freedom (DoF) of a free rigid body can be retrieved.

\rosso{Wireless detection of eye movements based on magnetometric measurements for eye-gesture estimation has been proposed as well, both in experimental animals (with surgically inserted magnets) \cite{schwarz_fsn_13, payne_ns_17, meng_fn_21} and humans (with magnetized contact lenses) \cite{tanwear_ieee_20}.}

This paper provides an analysis of the information that can be retrieved from the 5 DoF data
produced by a dipole tracker. We specifically address the case of a dipole eye-tracker, in which physiological constraints and/or predetermined maneuvers offer complementary information and make the tracking output data sufficient to fully determine the eye pose.

The paper is organized as follows: after a brief description of the developed instrumentation and of the tracking parameters obtained from the data analysis (Sec.\ref{sec:setup}), we describe in Sec.\ref{sec:gazeretrieval} several approaches that enable a complete eye-pose retrieval from the dipole pose information. In the same section some examples of the retrieved gaze trajectories are reported. In particular we discuss how the dipole-gaze misalignment may affect the estimated gaze trajectory and how physiological constraints can be taken into account to improve the accuracy. Conclusion and perspectives are drawn in Sec.\ref{sec:conclusion}.

\section{Setup}
\label{sec:setup}
\begin{figure*}[ht]
   \centering
      \includegraphics [angle=0, width= 0.45 \textwidth]  {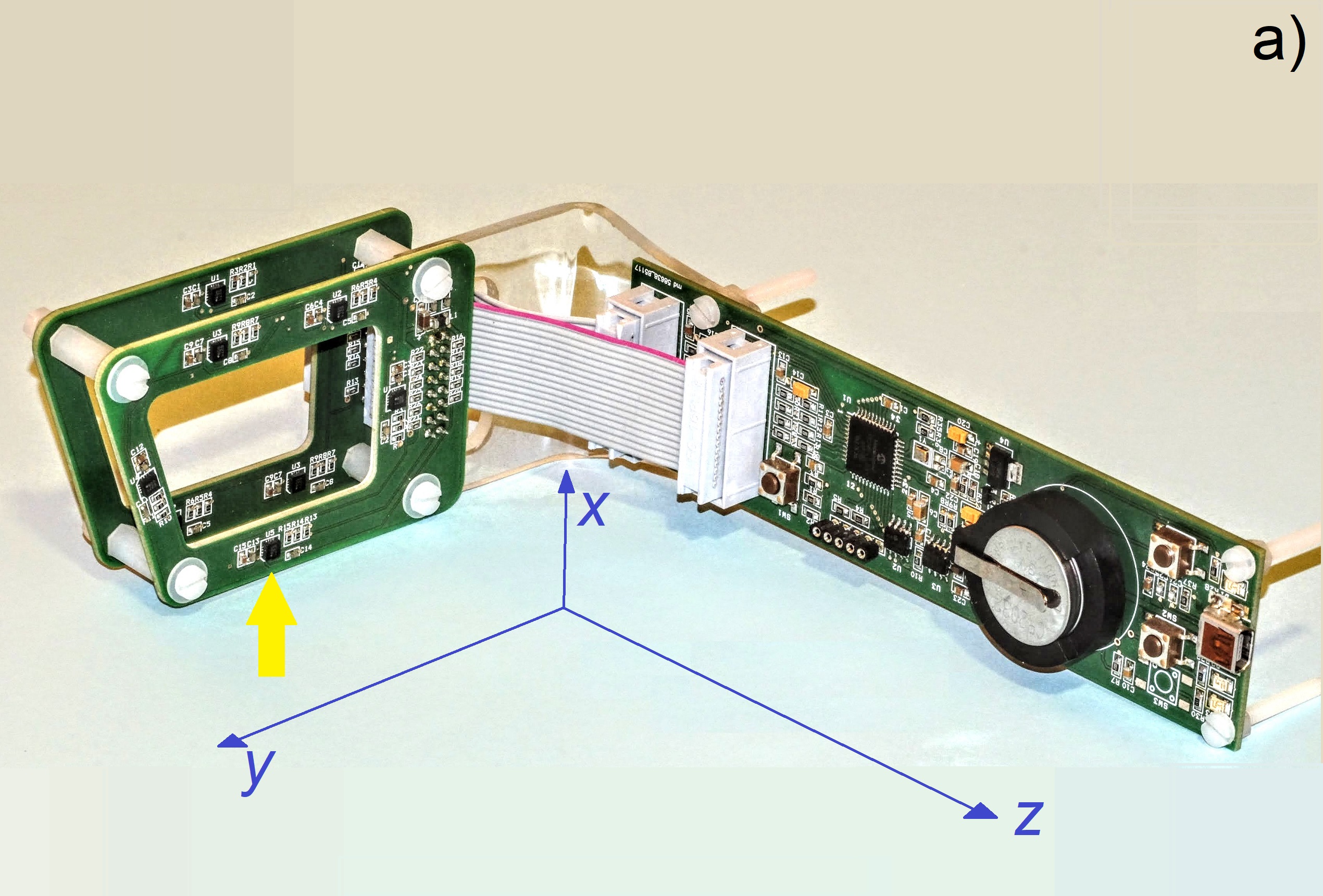}
      \includegraphics [angle=0, width= 0.45 \textwidth]  {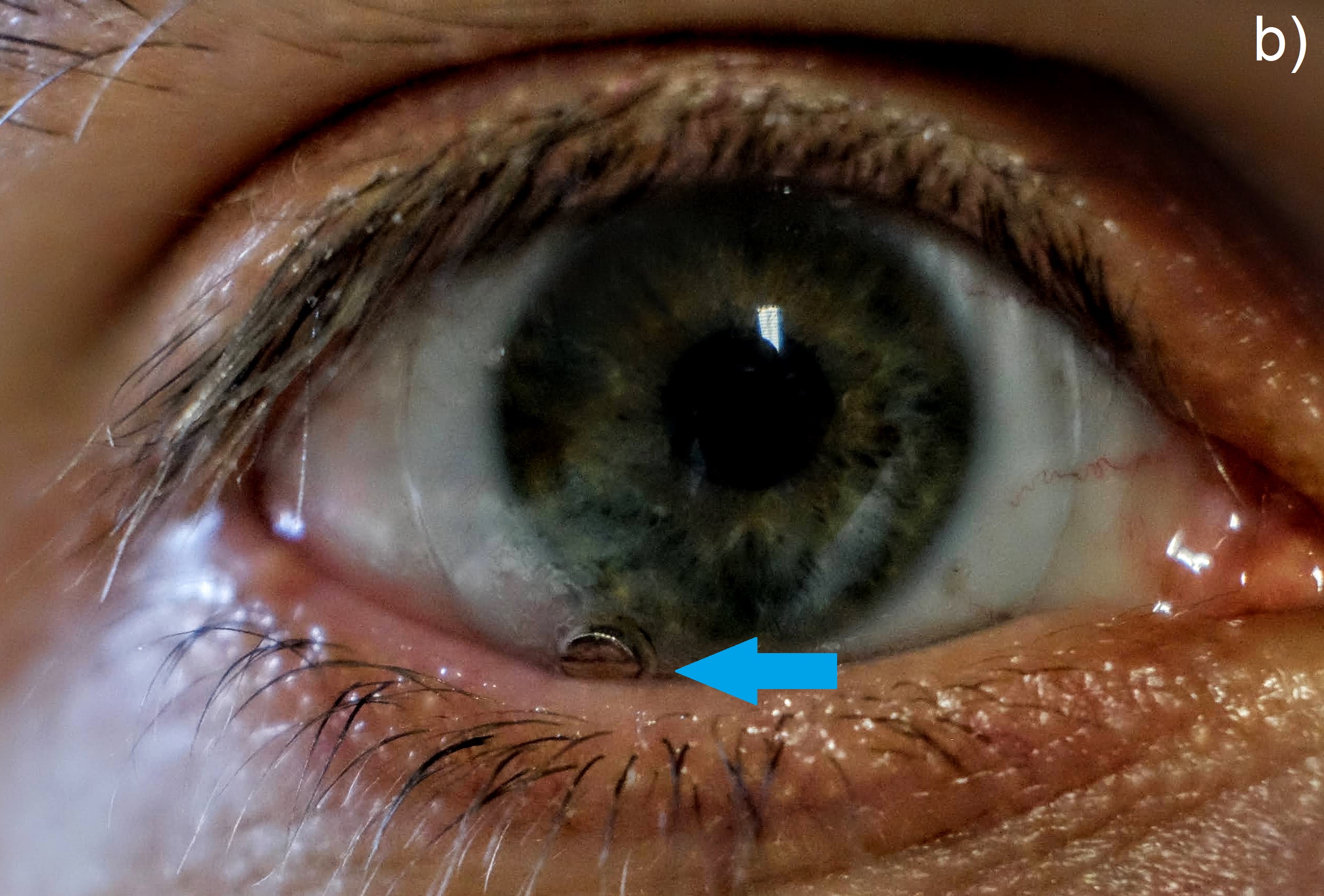}
      \caption{The sensor array --at the left side of image (a)-- contains eight three-axial magnetoresistive sensors (one of which is indicated by the yellow arrow) distributed on two parallel printed circuit boards (PCB) displaced by $\Delta z=16.6$~mm). A third PCB --at the right side of the image (a)-- hosts a microcontroller and other electronics necessary to store data during preliminary calibration procedures and to communicate with a personal computer \textit{via} a USB interface. The system is designed to localize a small magnet --pointed out by the blue arrow in the image (b)--, which is inserted in a scleral lens to be worn by the patient.\label{fig:sensorarray}}
\end{figure*}
\subsection{Sensors and hardware}

The hardware of the eye tracker is shown in Fig.\ref{fig:sensorarray} a. Its structure is extensively described in the Ref. \cite{biancalana_instrHW_21}. The core of the device is made of eight three-axial magnetoresistive sensors \cite{isentek8308} that generate a total of 24 data per measurement \rosso{(each datum corresponds to one component of the total magnetic field --Earth field plus dipole field-- measured in the position where the sensor is located)} . A microcontroller interfaces the sensors to a personal computer. The user may set sensitivity, acquisition rate and operation mode. Operation modes exist for  sensors' precalibration and for tracking operation, which includes data recording and/or immediate analysis and visualization.

The sensors are included in highly integrated circuits, which also contain preconditioning electronics and analog-to-digital converters (14 bit, up to 200~Sa/s). They communicate with the microcontroller \textit{via} eight independent I$^2$C buses, while the microcontroller is connected to PC \textit{via} a USB interface. This architecture helps accelerate the data transfer and makes possible to acquire synchronized measurements from all the sensors.
    
The magnetoresistances have a quite linear response, but their gains may differ from the nominal value causing an anisotropic response for each sensor and different responses from one sensor to the other. In addition, each magnetoresistance output is affected by a non-negligible offset. Thus a pre-calibration procedure is necessary, to equalize the gains and to identify and subtract the offsets \cite{biancalana_instrHW_21}.

\subsection{Data elaboration to extract tracking parameters}
\label{subsec:bestfit}
The main scope of the described instrumentation is to retrieve the eye orientation from the pose of a small magnet embedded in a scleral lens worn by the patient (Fig.\ref{fig:sensorarray} b). The position and the orientation of the target-magnet are inferred from a set of simultaneous field measurements performed in pre-assigned positions. The measured field  is modeled as a dipolar one generated by the magnet, superposed to an external one (the Earth field), which is assumed to be homogeneous over the volume occupied by the sensors.

The determination of the magnet pose (orientation and position) and intensity (dipole modulus), plus intensity and orientation of the environmental field constitute an inverse problem, which is solved by means of numeric tools. 

The size of the magnet (0.5 mm thick and 2 mm in diameter) is selected in such a way as to make its field on the sensors of the same order of magnitude as the Earth field (few tens of microtesla), which creates a good condition to accurately identify both the magnet pose and the environmental field.

Having equalized the readings to convert them in calibrated magnetometric data $\vec B^{\mathrm{(meas)}}_k$, solving the inverse problem requires to minimize the quantity:
\begin{equation}
    S= \sum_k \left |\vec B^{\mathrm{(meas)}}_k- B^{\mathrm{(mod)}}_k \right|^2
\end{equation}
being $k=0\dots7$ the sensor index, and
\begin{equation}
\begin{split}
\vec B^{\mathrm{(mod)}}_k (\vec r, \vec m)  = \frac{\mu_0}{4 \pi} \\
 \left ( 3\frac{[\vec m \cdot (\vec r^{(k)}-\vec r)](\vec r^{(k)}-\vec r) }{\left|\vec r^{(k)}-\vec r\right|^5}
-\frac{\vec m}{\left|\vec r^{(k)}-\vec r\right|^3}
\right)+\vec B_{geo},
\end{split}
\label{eq:modellodipolo}
\end{equation}

the  field expected on the basis of the dipole model, in the position $\vec r^{(k)}$ of the $k$th sensor, being $\vec m$ a magnetic dipole placed in the position $\vec r$. The  term $\vec B_{geo}$ accounts for the uniform environmental field. 
The positions $\{\vec r^{(k)}\}$ are referred to the sensor frame, which is rigidly connected to the patient's head.
The best fit procedure minimizes $S$ and outputs the estimates of $\vec r$, $\vec m$, $\vec B_{geo}$, for a total of nine tracking parameters. As discussed in Ref.\cite{biancalana_instrSW_21}, under some conditions, the number of fitting parameters can be reduced to eight after having determined the modulus of $\vec m$, while it is definitely inopportune to assume a fixed value for $|\vec B_{geo}|$, because the environmental field is commonly enough homogeneous on the volume of the array, but can vary significantly when the array is moved within the room where the system operates.

\rosso{We have investigated the reliability of ordinary best-fitting procedures obtaining encouraging results \cite{biancalana_instrSW_21}. In particular, we verified that a Levenberg-Marquardt minimization algorithm is sufficiently fast and  adequately robust with respect to the initial guess: there exists a volume larger than the region in which the magnet can move, such that any calculation starting from a dipole guess in that volume,  systematically converges to the correct solution, regardless of the  initially assigned orientation.}

\subsection{From tracking parameters to eye (and head) pose}
\label{subsec:param2eyegaze}
Once the minimization algorithm outputs the estimate of $\vec r$, $\vec m$ and $\vec B_{geo}$, the pose of eye and head can be inferred. While $\vec r$, $\vec m$ provide information of the eye pose with respect to the sensor array, the measure of the environmental field can be used as a three-dimensional compass to evaluate the absolute orientation of the sensor array, i.e. of the patient's head. 
Combining relative eye-gaze and head orientations enables
the determination of the absolute gaze \cite{bellizzi_rsi_22} \rosso{direction. Notice that the system is non-sensitive to head translation, thus the retrieval of the gaze direction is not sufficient to identify a fixation point at a finite distance.}

In addition, the comparison of eye and head angular motions can  be used
to evaluate the eye actuation in response to head rotation, such as in measurements of vestibulo-ocular reflex (VOR)
\cite{crawford_jn_91} (see also Sec.\ref{subsec:oneaxis}). However, this paper focuses on the analysis of the  eye pose relative to the head, and we are not considering further the retrieval of head pose from $\vec B_{geo}$.

The orientation of $\vec m$ is related to eye rotations, but it does not provide  the gaze direction directly, unless the magnetic dipole is parallel to the optical axis of the eye (let the latter be $\hat e$). Eye rotations around the $\vec m$ direction would not be revealed. This has a twofold consequence: if $\vec m \parallel \hat e$, torsional angular displacements are not detected; if $\vec m \nparallel \hat e$, some \textit{extra assumptions} are necessary to infer $\hat e$ from $\vec m$, and, in particular, rotations around $\vec m$ would affect $\hat e$ without being detected.

It is worth recalling that besides $\vec m$, the best-fit procedure outputs the  magnet position $\vec r$, which provides additional information about eye motion. However, the position co-ordinates are known with respect to the sensor reference frame and not to the eye center, thus the eye gaze cannot be retrieved directly from $\vec r$. Despite this feature, the partially redundant information given by $\vec r$ could be used in conjunction with that extracted from $\vec m$ to improve or to complete the eye pose reconstruction, as it will be discussed in Sec.\ref{subsec:r2gaze}.

\section{Gaze retrieval}
\label{sec:gazeretrieval}

In this section we examine some procedures that can be followed to retrieve the eye gaze with respect to the sensor frame from the tracker output,  more specifically from the estimates of $\vec m$ and $\vec r$. 
Eye translations, that can in principle be studied with an adequate analysis of $\vec r$ 
will not be taken into consideration, and we
will focus on eye-ball rotations, modelling  the eye as a rigid sphere rotating about its (fixed) center.
Although mathematically unnecessary, this will turn in a useful simplification of the following $\vec m$-related discussion with no further hypothesis or loss of information.
Besides, the fixed-center rigid sphere model applies to the $\vec r$ information discussion in the realistic approximation of very small displacements of the eye-ball center compared to the eye radius. 

It has to be pointed out that the pose of a rigid body freely rotating about an assigned point requires three angular parameters to be fully defined. In other terms, generally speaking, the faced problem 
 has 3 degrees of freedom (DoF). 
The modulus of the magnetic dipole is nominally constant, and the tracking uses its direction $\hat m$ only, thus the analysis of the unitary vector $\hat m$ provides only bidimensional information, e.g. the zenithal and azimuthal angles, when described in spherical co-ordinates. These 2D data are not sufficient to determine the 3 DoF orientation of the eye-ball: it is necessary to consider reduced-DoF problems to achieve complete pose determination.

\subsection{Small-angle and near-front sight approximation }

\label{subsec:smallrotations}
A simple case is obtained in the hypothesis that torsional movements can be neglected and that the angle between the gaze and the front direction ($\hat z$  in the notation of our setup, see Fig.\ref{fig:sensorarray}) is maintained small.

In this hypothesis the current gaze  \rosso{$\hat e_i$} is always nearly parallel to $\hat z$  and it can be expressed as the rotation of the front gaze direction $\hat e_0$, \rosso{$\hat e_i= \mathbf{R} \hat e_0$}, being
\begin{equation}
\mathbf{R}= 
\begin{pmatrix} 
	1          &     0      & \phi_y    \\
	0          &     1      & \phi_x    \\
	-\phi_y    &  -\phi_x   & 1         \\
\end{pmatrix}
\end{equation}
the first-order approximation of a rotation matrix whose parameters $\phi_x, \phi_y$ 
represent small rotation angles around the $x$ and the $y$ axes respectively. The  same rotation applies to the magnetic moment $\vec m_0$, as it is rigidly connected to the eye-ball. Thus \rosso{$\hat m_i=\mathbf{R} \hat m_0$}, and the two  angles $\phi_x, \phi_y$ can  be retrieved from this last relation.

In summary, here $\hat e_0$ is the front sight gaze, $\hat m_0$ the corresponding orientation of the dipole and \rosso{$\hat e_i$} the gaze at the $i$th measurement, to be inferred from the estimated dipole \rosso{$\vec m_i$}. In this approach, $\hat e_0$ is assumed to coincide with $ \hat z$, and $\hat m_0$ should be known from an independent characterization of the magnet in the scleral lens, or tentatively assigned as it will be explained below.

\subsection{Larger rotations about the front ($\hat z$) direction}
\label{subsec:largerotations}
A more complex case occurs when the first-order approximation is not applicable. If so, even with the simplifying hypothesis $\hat e_0 \parallel \hat z$, the eyeball rotation cannot be, in general, resolved as a sequence of two rotations around arbitrary axes. As we will see in the following, further information is needed in this case, like a known rotation axis or a model for the eye movements.
As an intuitive proof of this, consider the two-angle  rotations around the pre-set Cartesian axes $x$ and $y$ defined by $\mathbf{R}_{yx}=\mathbf{R}_y \mathbf{R}_x$ or $\mathbf{R}_{xy}=\mathbf{R}_x \mathbf{R}_y$.  
These have the general form
\begin{equation}
\begin{split}
\mathbf{R}_{yx} & =
\begin{pmatrix} 
	\cos \phi_y          &     -\sin \phi_x \sin\phi_y      & \cos \phi_x \sin\phi_y    \\
	0                    &     \cos \phi_x                  & \sin \phi_x    \\
	-\sin \phi_y         &  - \sin \phi_x \cos \phi_y       & \cos \phi_x \cos \phi_y        
\end{pmatrix},\\
\mathbf{R}_{xy} & =
\begin{pmatrix} 
	\cos {\phi_y}'             &     0                            & \sin{\phi_y}'    \\
	-\sin {\phi_x}' \sin {\phi_y}' &     \cos {\phi_x}'                  & \sin {\phi_x}'   \cos {\phi_y}' \\
	-\cos {\phi_x}' \sin{\phi_y}' &  - \sin {\phi_x}'        & \cos {\phi_x}' \cos {\phi_y}'         \\
\end{pmatrix},
\end{split}
\label{eq:RxRy}
\end{equation}
and both converge to the above operator (3) in the small angle approximation.
On the other hand, they represent intrinsically different operators for large angles\footnote{Indeed the two operators coincide if $\phi_x={\phi_x}'=0$ or $\phi_y={\phi_y}'=0$, that is if one of the two $x$ or $y$ axes happens to be the unique rotation axis. This has a practical utility that will be discussed in subsection \ref{subsec:oneaxis}}.

Indeed, the two angular data extracted from the direction of $\hat m$ are not sufficient to determine uniquely the 3-DoF  configuration, and the selection of a particular rotation order corresponds to arbitrarily assign the missing information. Such an approach allows for producing tentative
eye gaze trajectories, but the solution is not stringent and diverse choices results in diverse and variously distorted trajectories. Nevertheless, also with this under-specified 
analysis, the main features of the gaze trajectories can be identified, which can be satisfactory in some eye-tracking applications.

\begin{figure*}[ht]
\centering
\includegraphics[angle=0, width=0.4 \textwidth]{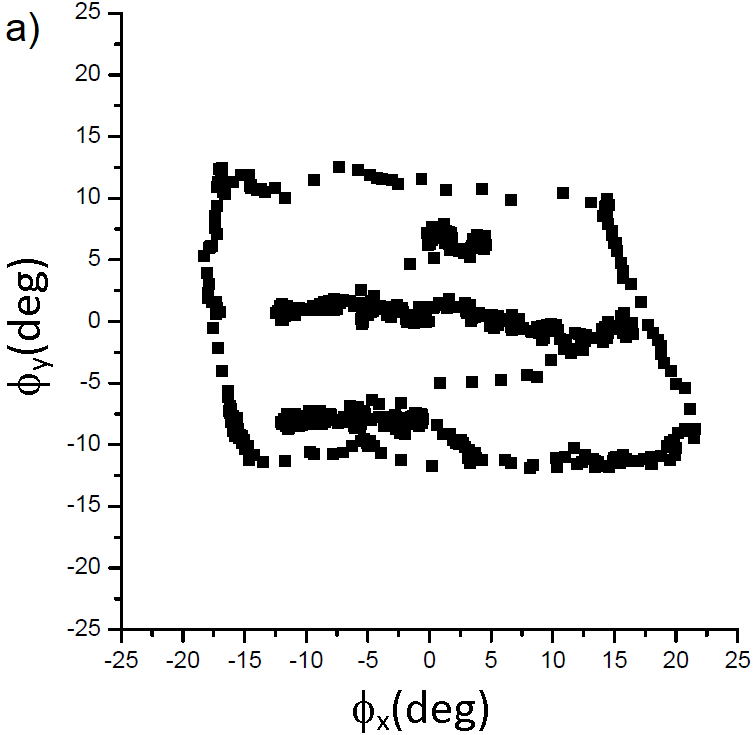}
\includegraphics[angle=0, width=0.4 \textwidth]{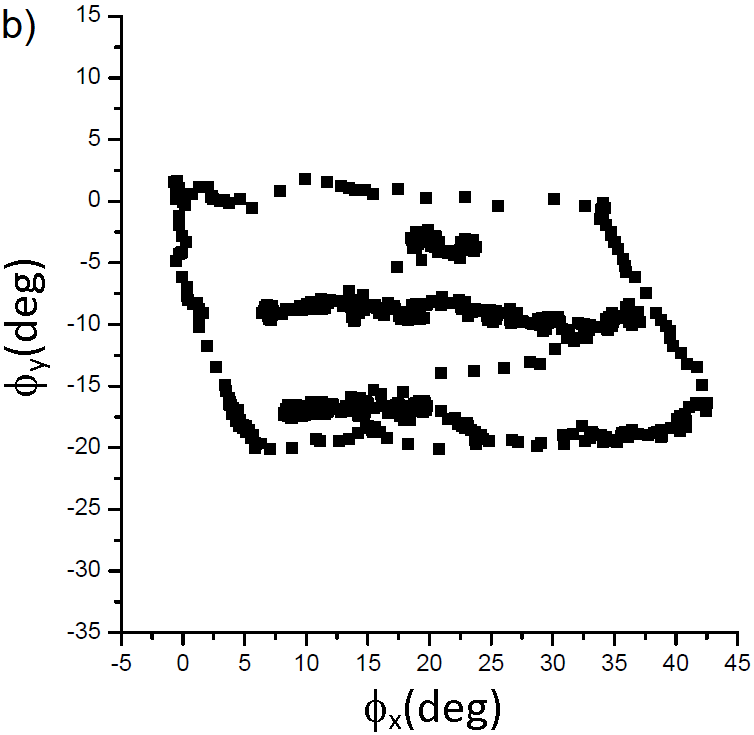} 
\includegraphics[angle=0, width=0.4 \textwidth]{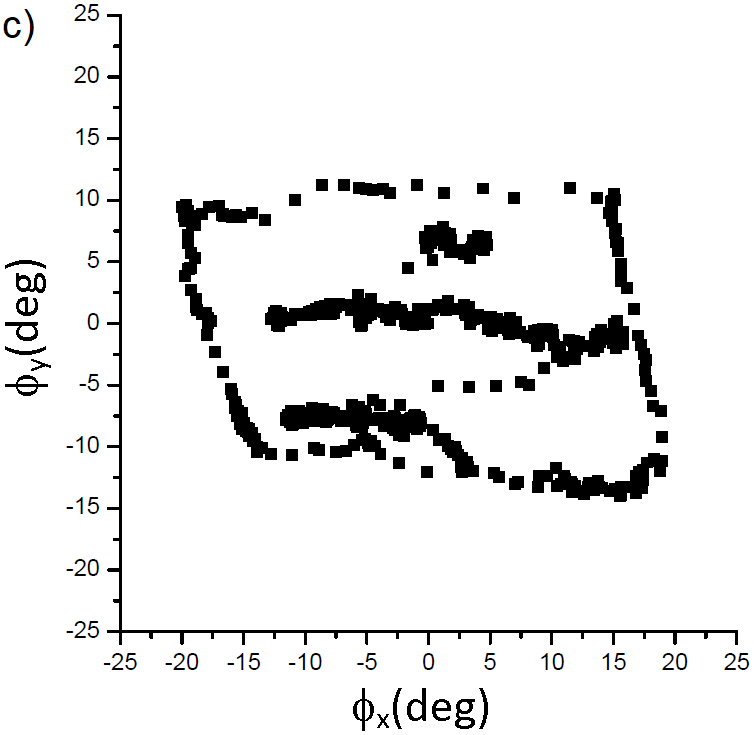} 
\includegraphics[angle=0, width=0.4 \textwidth]{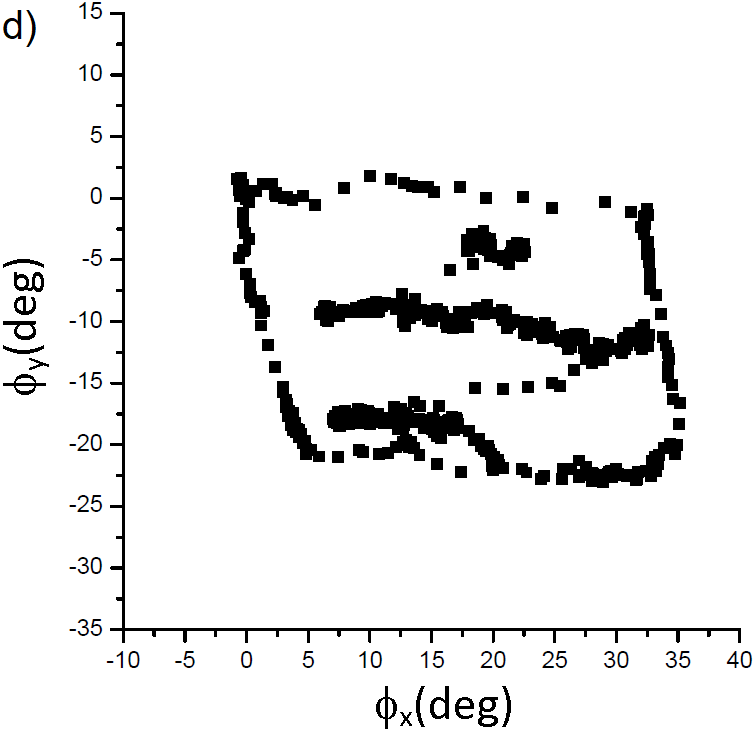} 
\caption{Gaze trajectories reconstructed on the basis of the approach described in Sec.\ref{subsec:largerotations}. The subject was sitting about 110 cm away from a monitor, whose frame is 40 cm $\times$ 71.5 cm in size. He was requested to read a three-line text and then to follow the monitor frame (see also Fig.\ref{fig:ListingInVivo}). The upper plots (a,b) are obtained using $\mathbf{R}=\mathbf{R}_y\mathbf{R}_x$, while the lower ones (c, d) are obtained with $\mathbf{R}=\mathbf{R}_x\mathbf{R}_y$. 
The front sight direction (0,0) is not known and is in turn assigned to a central point of the screen (left plots, a, c), or to the upper-left corner (right plots, b, d).
The small displacements of the trajectory are well tracked in all the cases, but evident image distortions occur over the large distances, i.e. when  large rotations are involved. This kind of distortion arises independently on the assignment of the front-sight direction.\label{fig:NoListingInVivo}}
\end{figure*}

As an example, in Fig.\ref{fig:NoListingInVivo} we represent tracking results obtained in a in-vivo experiment, using the $\mathbf{R}_{xy}$ and $\mathbf{R}_{yx}$ operators alternatively. The subject was requested to read a short text (title, a complete line, a half-line) on a monitor and then to follow the monitor frame. 
The four represented trajectories are obtained from the same dataset, with an analysis of the dipole orientation based upon the two operators, respectively. In this elaboration, one selects (tentatively) the $\hat m_0$ orientation, i.e. the trajectory point to which the $(\phi_x=0, \phi_y=0)$ co-ordinates are being assigned in such a way to obtain as  straight lines and right angles as possible.
As one can see, in all cases features like  text line distribution and  small displacements around the running line of sight are well tracked, while distortions can't be canceled and affect long segments that should be straight and  angles at the monitor corners that should be right.

The subject was sitting in front of the monitor so that $\hat e_0$ is expected to coincide with (to be in the proximity of) the measurement tracked at the monitor center. Two of the trajectories --(a) and (c)-- are obtained in this hypothesis, assigning $\hat e_0$ to the central point of the figure (thus selecting the corresponding $\hat m$ as $\hat m_0$). The other trajectories --(b) and (d)-- are instead obtained with a wrong assignment ($\hat e_0$ in coincidence of the upper-left corner, about 20 degree away from the nominal direction). As can be seen, the main features of the trajectory are well reproduced also in this case, which demonstrates robustness with respect to the $(\hat e_0, \hat m_0 )$ assignment.

Is worth reminding that in principle, also the $\hat e_0$ could be at some angle from $\hat z$, however this angle amounts at less then 10° and it would cause minor changes to the plots shown in Fig.\ref{fig:NoListingInVivo}.

\subsection{Large rotations about an assigned axis}
\label{subsec:oneaxis}

The issue related 
to the non commutativity of  finite 3D rotation is overcome in a preeminent application of the eye-tracking systems. In VOR measurements, the head of the patient is rotated around a predetermined axis, which  is usually vertical (pitch) or horizontal (yaw),  the expected (ideal) response of the eye is an opposite (compensating) rotation around the same axis, and this is the quantity to be estimated and analyzed.

In this case, the data are to be interpreted in terms of a single rotation angle around one axis. That axis can be assumed to coincide with the nominal one that is \textit{approximately} either the $\hat x$ or the $\hat y$ direction defined in Fig.\ref{fig:sensorarray} in the cases of pitch or yaw, respectively.

Improved accuracy is achieved if the actual rotation axis is identified on the basis of the $\vec B_{geo}$ measurement set \{$\vec B_i$\} ($i=1,\dots,N$)  recorded during the maneuver. For instance, one can find the common plane intercepting the measured $\vec B_i $ vectors (with a mean square distance minimization). The perpendicular direction to the plane, $\hat u$, provides an estimate of the rotation axis direction \cite{pearson_js_901, schomaker_ac_59}.
Subsequently, both the head and eye rotations can be evaluated around the determined direction. The eye rotation angle can be estimated from two measured values $\vec m_i$ ($i=1,2$) as
\begin{equation}
    \phi_{u,1}-\phi_{u,2}=\arcsin{ \frac{|\vec m_{1, \perp } \times \vec m_{2, \perp }|}{|m_{1, \perp }||m_{2, \perp }|}}
    \label{eq:arcsin}
\end{equation}
being
$\vec m_{i, \perp }= \vec m_i -(\vec m_i \cdot \hat u) \hat u $,
and similarly --using $\vec B_{geo}$-- for the head. 

This perfecting produces just a slight enhancement of the accuracy, if the misalignment between ideal and actual rotation axes is small, i.e. $\hat u \approx \hat x$ or $\hat u \approx \hat y$. 
Whatever the actual rotation axis is, in this application eye and head rotations never exceed $\pi/2$, which makes unnecessary to refine eq.\ref{eq:arcsin} for out-of-range arcsine issues\footnote{\rosso{An alternative method to determine the rotation axis is fitting a 3D circle on the {$\vec B_i$}, as $\vec B$ moves nominally on a circle at constant zenithal angle with respect to the rotation axis \cite{fisher_url_17}. 
To this end, we have implemented a procedure that:
i) applies a rotation matrix parametrized with a couple of Euler angles $(\alpha, \beta)$ to a set \{$\vec B_i$\} or \{$\vec m_i$\};
 ii) represents the data in spherical co-ordinates;
iii) determines the couple $(\alpha, \beta)$ as that minimizing the standard deviation of the zenithal angles. 
}}.

\rosso{It is worth noting that assuming that the axes of rotation of the head and eye coincide is a popular but coarse approximation. Indeed,} considering the case of a generic rotation, 
the VOR gain is not isotropic \cite{aw_jn_96, pogson_jn_19}, so that --also in the simplified assumption of a linear and non-delayed response-- the VOR gain should be expressed tensorially. E.g., focusing on the angular velocities, the relationship between eye and head  speed would be expressed as $\vec \omega_e=\mathbf{G} \vec \omega_h$, where $\mathbf{G}$ is a $3 \times 3$ matrix. $\mathbf{G}$ is expected to be a non-diagonal matrix, implying that eye and head rotation axes are generally not  parallel. A correct and complete analysis of the VOR gain involves the determination of all the elements $\mathbf{G}_{ij}$, which (at least in principle) can  be obtained from the above described data. 

As an additional remark, let's point out 
that our prototype contains only magnetometric sensors, however improved setups could be designed to include gyroscopic detectors. In this case, additional and complementary estimates of the head rotations would be available, with improved accuracy in VOR gain estimations.

\subsection{Physiological constraints}
\label{subsec:listing}

\begin{figure*}[ht]
        \centering \includegraphics [angle=270, width=0.7\textwidth] {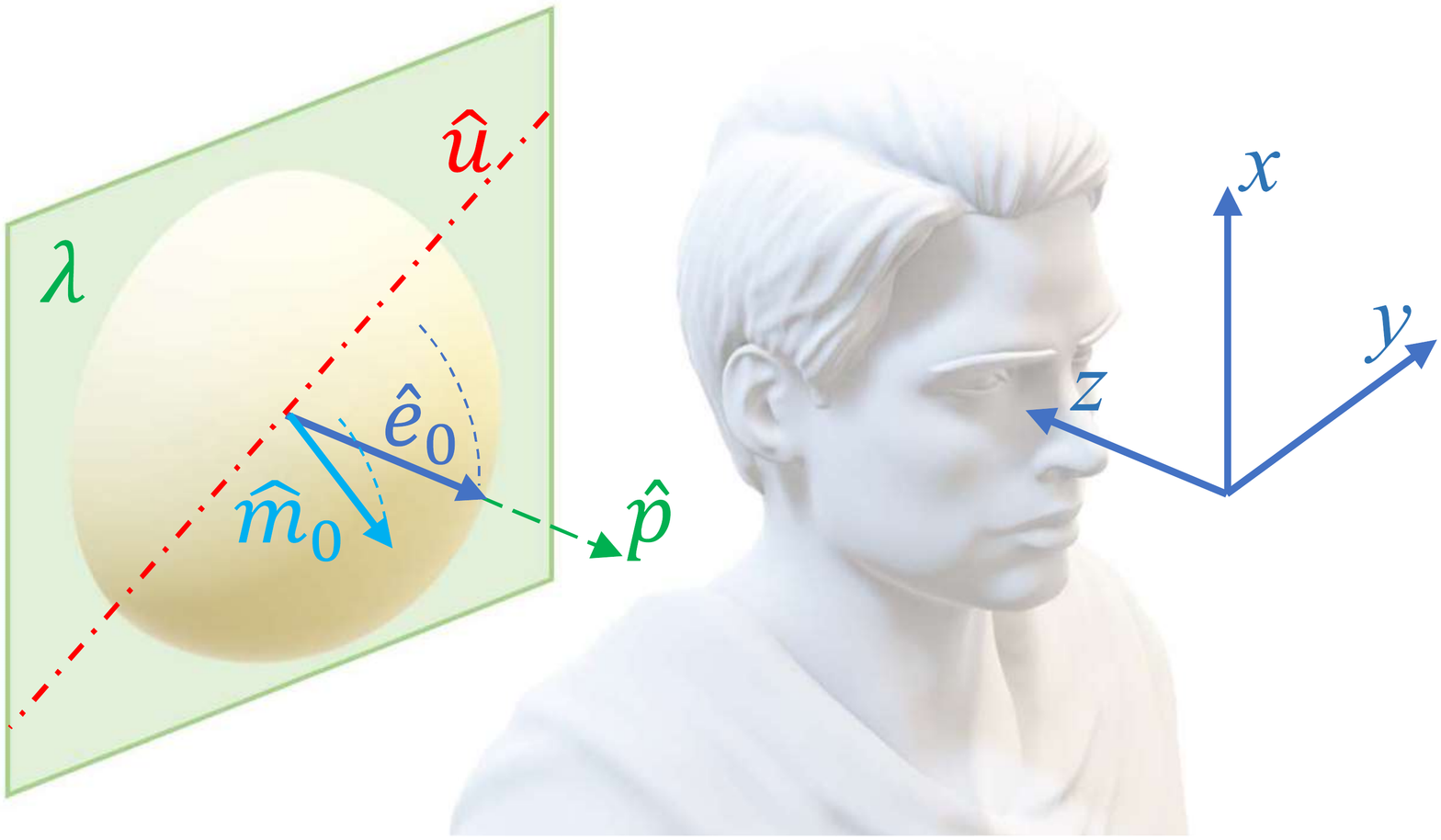}
  \caption{Definition of the axes and geometry of the Listing's model. The primary orientation $\hat e_0$ (along the normal $\hat p$ to the Listing's plane $\lambda$) is nearly parallel to $\hat z$. In our setup, when the eye is in its primary orientation the magnetic dipole is generically oriented ($\hat m_0$). When the gaze varies, both vectors $\vec m$ and $\hat e$ undergo rotations (dashed arcs) obeying the Listing's law: each accessible configuration is such that it could be reached from the primary orientation with one rotation around an axis laying on $\lambda$. In this figure, the direction $\hat u$ embodies one of those axes.
   }
   \label{fig:definizioniListing}
\end{figure*}

The arbitrariness of the rotation matrix introduced in Sec.\ref{subsec:largerotations} (eq.\ref{eq:RxRy}) can be avoided if the DoF reduction is performed on the basis of physiological constraints of the eye. 
Torsional movements of the eye are strongly depressed and in a good approximation the eye movements respect a 1D constraint enunciated in the 19th century and named Donders' law. 
Abundant literature is available on this subject. The reader may find a useful introduction to the matter in Ref.\cite{wong_so_04} and more mathematical details in Ref.\cite{haslwanter_vis_94}. 
Donders' law states that each direction of the gaze corresponds to a single orientation of the eye, regardless of the trajectory that led to the corresponding eye pose. In other words, each direction of view corresponds to a unique eye configuration and torsional movements around the current line of sight do not occur or can be neglected.

As far as the Donders' law is respected, the eye movements have only two DoF, hence it is possible to search for a biunique relationship between $\hat m$ and $\hat e$. A second phenomenological law that rules the eye movement is known as Listing's law. It suggests a good parametrization to describe these two-DoF configurations allowed by the Donders' law. 

According to the Listing's law, there exists a direction (primary orientation)  ($\hat p$) such that any eye configuration can be described as if reached moving from that primary orientation by means of a single rotation around an axis perpendicular to $\hat p$, as sketched in Fig.\ref{fig:definizioniListing}.
In other terms, making reference to a spherical co-ordinate systems having the polar axis along $\hat p$, all the eye configurations can be reached with one opportune zenithal rotation from the primary orientation: the rotation axes lay on the equatorial plane of that co-ordinate system, the so-called Listing's plane. When the head is erect and the eye is at the primary orientation, the gaze is approximately straight ahead, thus the Listing's plane is not far from being vertical. 

Summarizing, the Listing's law reduces the DoF of eye configuration to two: each eye configuration can be considered as the result of a rotation by a given zenithal angle around an axis $\hat u$ that lays on the Listing's plane and has a given orientation (azimuthal angle) on that plane. The data elaboration needed to retrieve $\hat e_i$  from $\hat m_i$ consists in the determination of those two angles from the current orientation of the dipole ($\hat m_i$) and its orientation ($\hat m_0$)  when the gaze is along the primary direction. This problem is focused in next Sec. \ref{subsec:mylisting}.
\subsection{Applying the Listing's law to the present case}
\label{subsec:mylisting}
\begin{figure*}[ht]
   \centering
        \includegraphics [angle=270, width= 0.7 \textwidth] {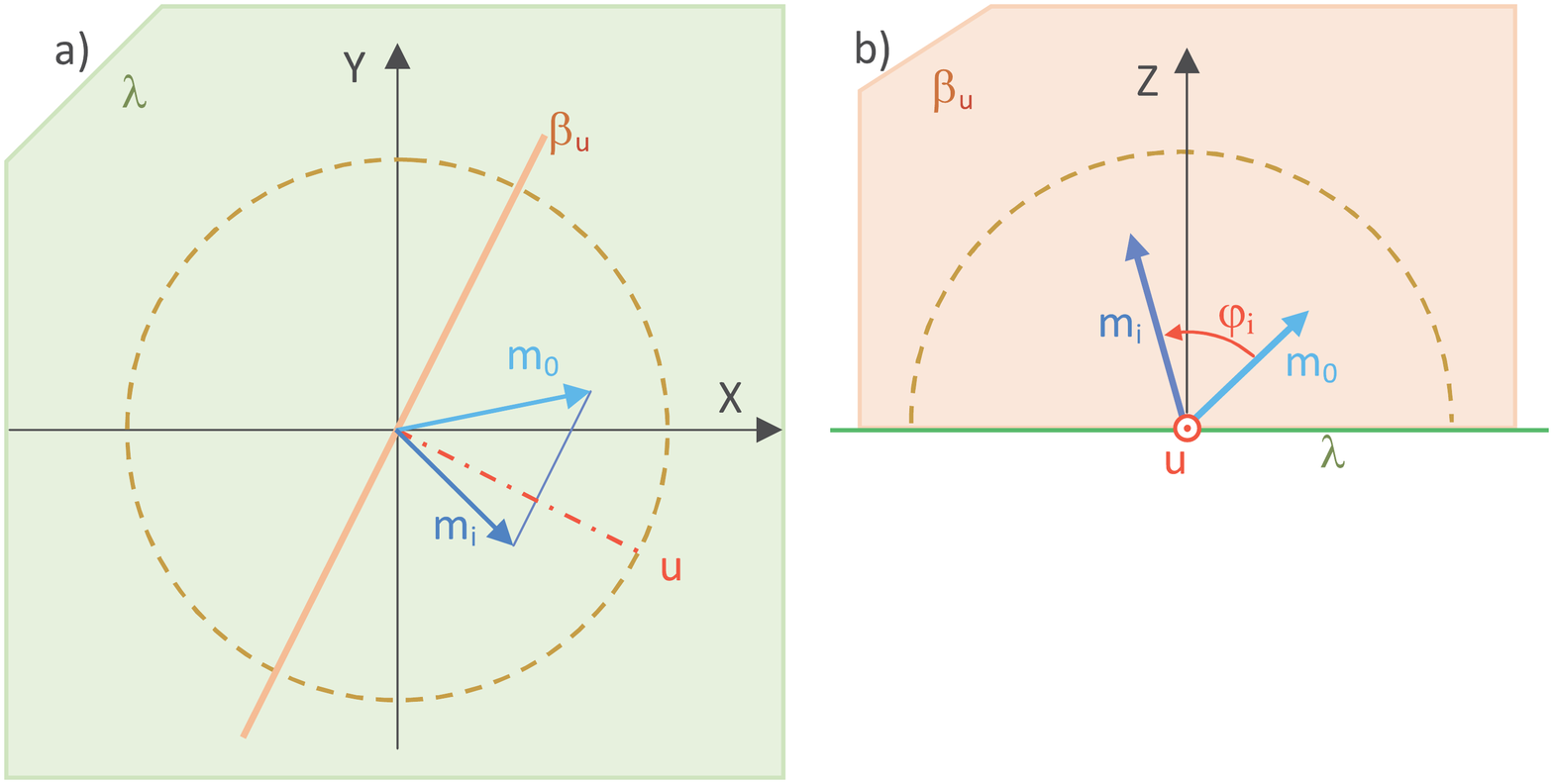}
        \caption{ In the rotated system $XYZ$, the Listing's plane $\lambda$ is the $XY$ plane. In (a) the projections of $\hat m_0$ and $\hat m_i$ on that plane are represented. The rotation axis $\hat u$ lays on that plane and is identified as the normal to the difference between the projections of $\hat m_0$ and $\hat m_i$. The direction of $\hat u$, together with the $Z$ axis, identifies the plane $\beta_u$.  In (b) the projections of $\hat m_0$ and $\hat m_i$ on $\beta_u$ are shown: $\beta_u$ is the plane where  the rotation angle $\varphi_i$ around $\hat u$ can be evaluated.
        }
  \label{fig:listingplane}
\end{figure*}

In our implementation, $\hat p$ is approximately antiparallel to $\hat z$. Let  $\hat p$ be known or tentatively assigned (see below).  First of all we refer the tracking data to a rotated coordinate system $XYZ$ such that $\hat p =- \hat Z$. To this end, we determine the Rodrigues' matrix \cite{rodrigues_jmpa_40, cheng_jam_89} that leads $\hat z$ to coincide with $\hat Z=-\hat p$ and we use this matrix to represent the measurements {$\vec m_i$} in these new co-ordinates $XYZ$, where the Listing's plane is the $XY$ plane\footnote{Note that the reference axis $X$ can be arbitrarily chosen on the Listing's plane, since all data of interest are related to angular displacements (and not to absolute directions).}.

Let $\hat m_0$ and $\hat e_0$ be the magnet and the eye directions corresponding to the primary orientation:  by definition, $\hat e_0= -
\hat Z$, while $\hat m_0$ must be known or tentatively assigned.
The task is now to determine the orientation $\hat e_i$ corresponding to the measured $\hat m_i$ on the basis that $\hat e_i=R_{\varphi_i, \hat u_i}\hat e_0$ and $\hat m_i=R_{ \varphi_i, \hat u_i}\hat m_0$, where $R_{ \varphi_i, \hat u_i}$ is a rotation by an angle $\theta$ around the direction $\hat u$ that lies on the $XY$ plane (again, the Rodrigues' formula can be used to determine the corresponding matrix).

$R_{ \varphi_i, \hat u_i}$ can be uniquely determined from $\hat m_0$ $\hat m_i$ as follows. The rotation axis $\hat u = (u_X, u_Y, 0)$ is perpendicular to the projection of $\Delta \hat m=\hat m_0 - \hat m_i$ on the $XY$ plane: $\hat u \cdot (m_{0X}-m_{iX}, m_{0Y}-m_{iY}, 0)=0$.
Now consider the projections of $\hat m_0$ and $\hat m_i$ on the plane perpendicular to $\hat u$: $\varphi_i$ the angle between those projections can be evaluated with the method described by eq.\ref{eq:arcsin}. The retrieved $\hat u$ and $\phi_i$ give the axis and the rotation angle, respectively, in the Listing's model of the eye. 

\begin{figure*}[ht]
\centering
\includegraphics[angle=0, width=0.4\textwidth]{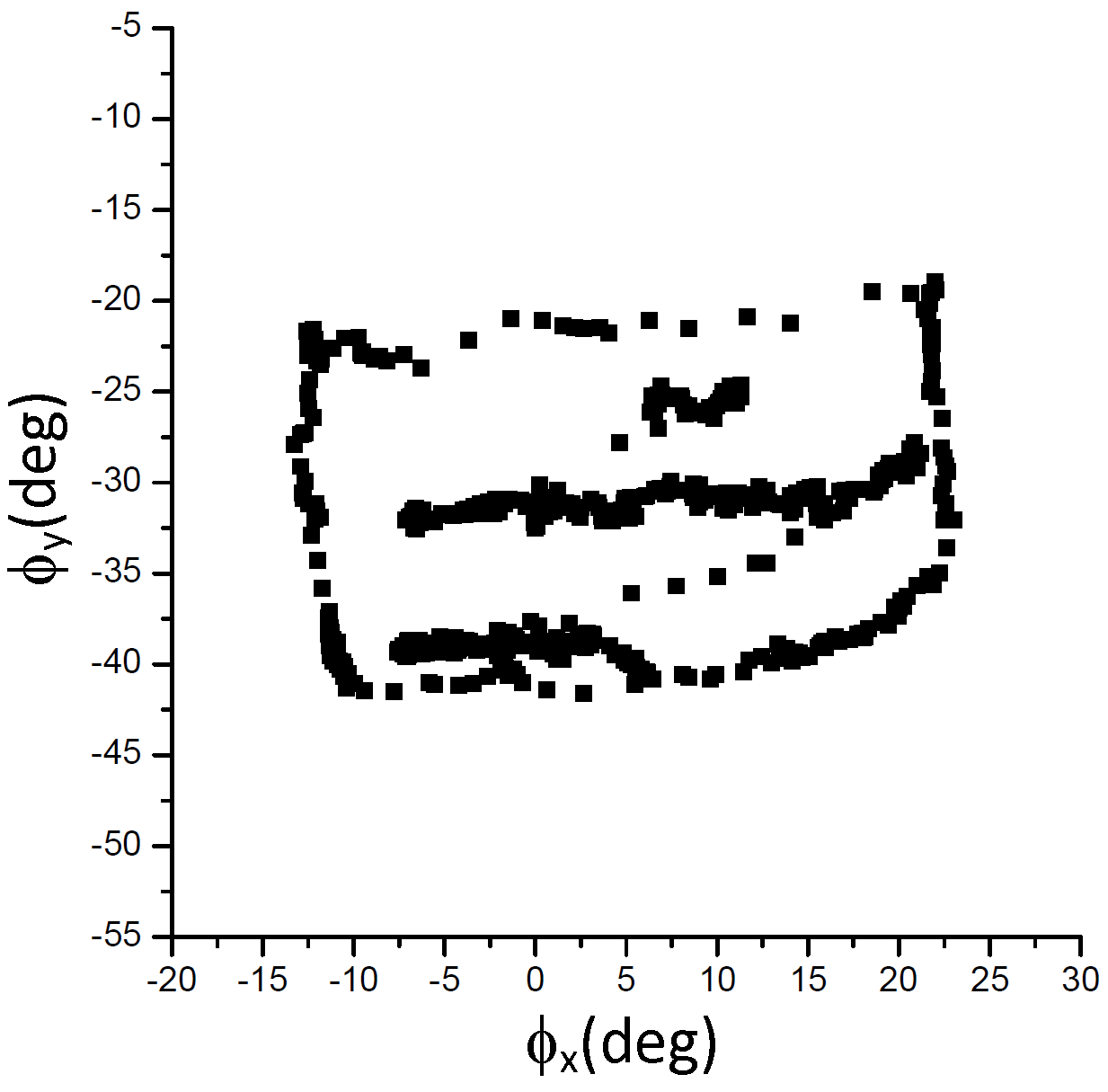}
\includegraphics[angle=0, width=0.4\textwidth]{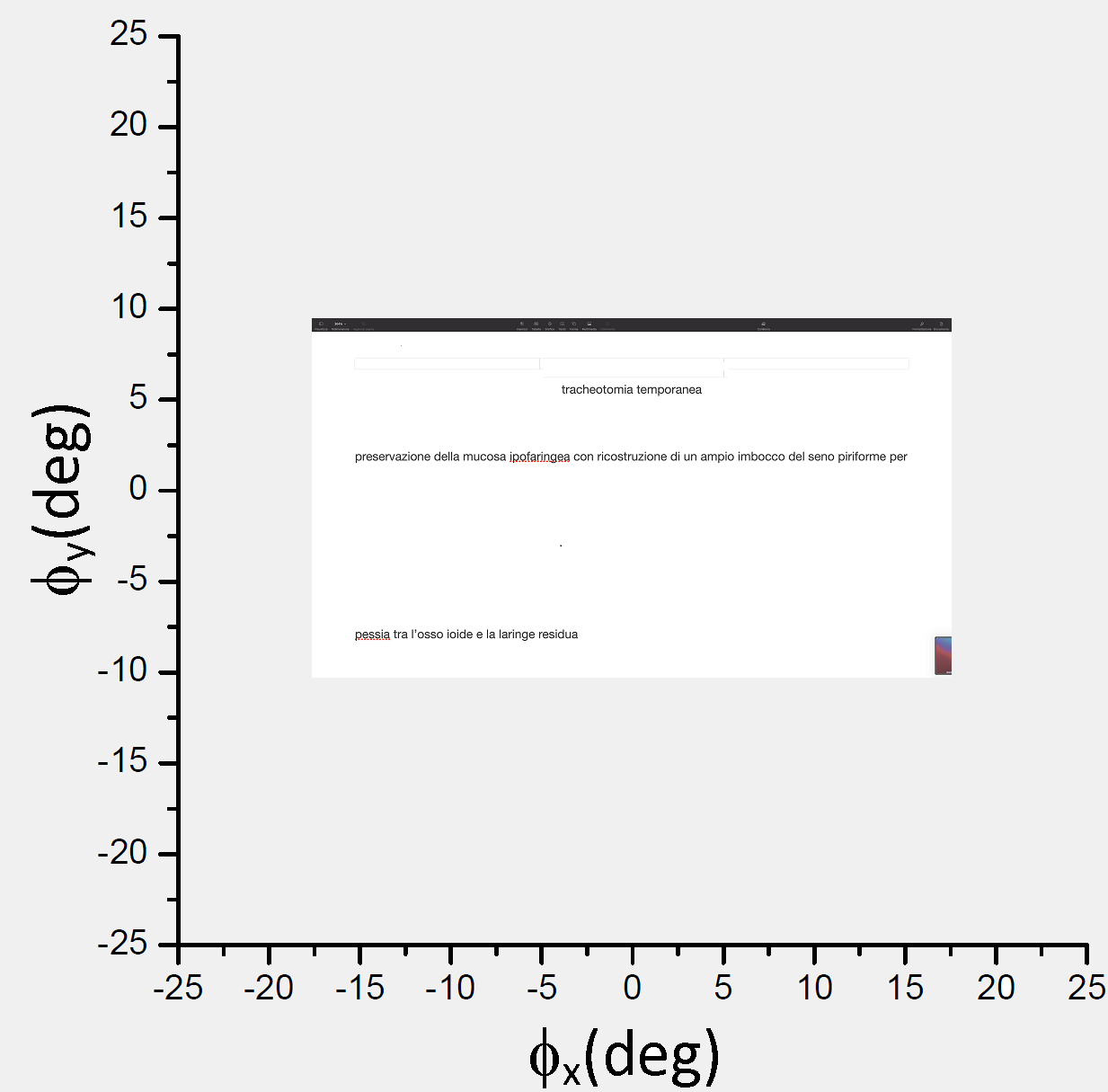}
  \caption{The same data used to produce the gaze trajectories represented in Fig. \ref{fig:NoListingInVivo} are here elaborated on the basis of the Listing's model. With an appropriate selection of the $\hat m_0$ and $\hat p$ directions, the whole data set fits with much lower distortion onto a shape that reproduces the text and the monitor frame observed by the subject (sketched in the right figure). 
  } \label{fig:ListingInVivo}
\end{figure*}

When $\hat p$ and/or $\hat m_0$ are not known \textit{a priori}, they can be tentatively assigned and their appropriateness can be evaluated on the basis of the distortion of the reconstructed gaze trajectory corresponding to regular and known shapes. This is the case of the trajectory represented in Fig.\ref{fig:ListingInVivo}. The same recording used for Fig.\ref{fig:NoListingInVivo} is here processed according to the above described Listing's model procedure. A direct comparison with the trajectories shown in Fig.\ref{fig:NoListingInVivo} puts in evidence a clear reduction of distortion effects. 
On the basis of this observation, we conclude that
the Listing's law makes it possible to develop a good procedure to accurately retrieve the eye-gaze, despite the $\hat m - \hat e$ misalignment. 

\begin{figure*}[ht]
\centering
\includegraphics[angle=-90, width=0.6\textwidth]{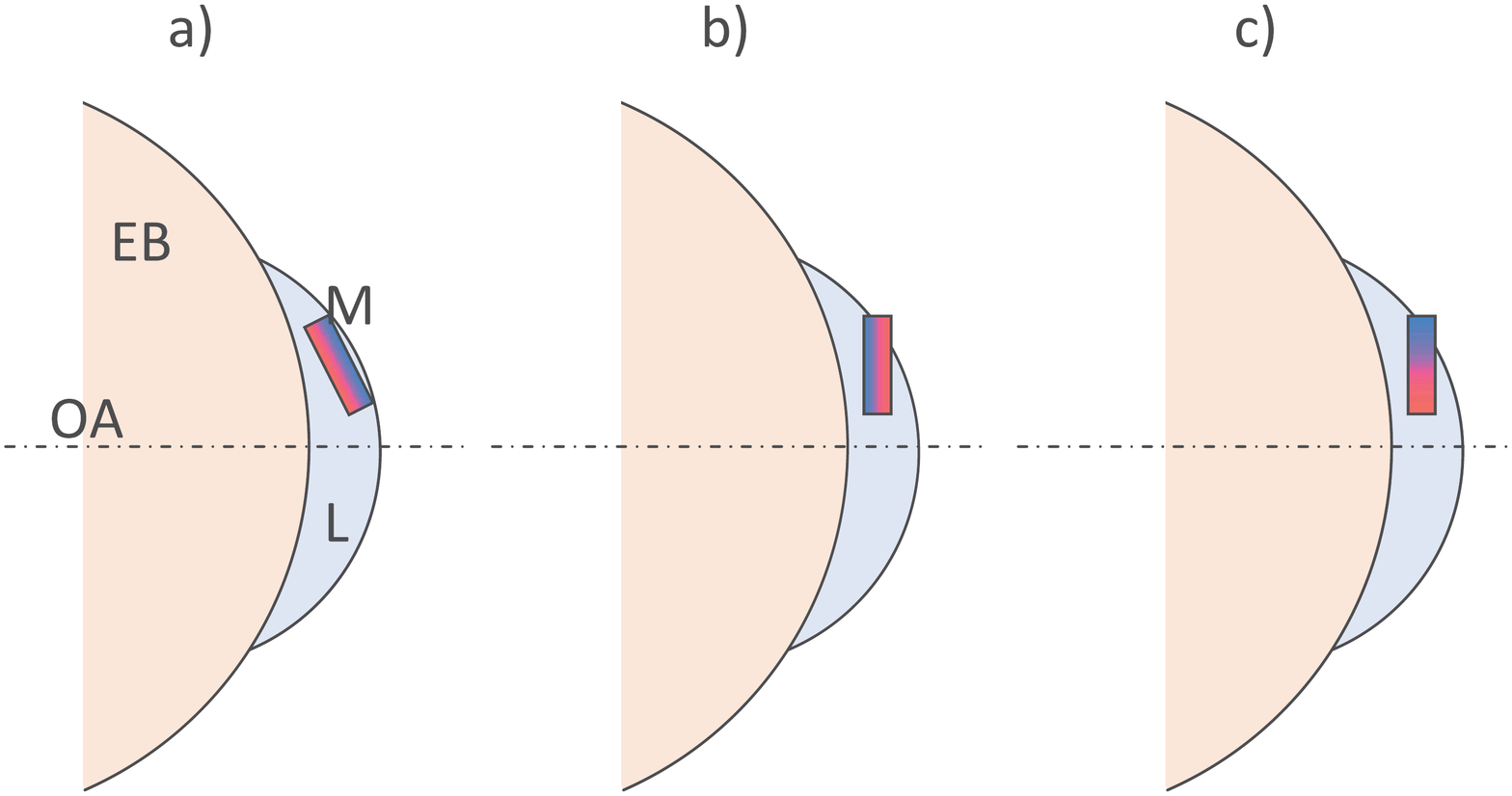}
  \caption{Schematic representation (not to scale) of an eye-ball (EB) (OA is its optical axis) wearing a lens (L) with a magnetized disk (M) embedded. The red-blue colors represent the M magnetization. In the current implementation (a), the disk is axially magnetized and is inserted tangentially to the lens, a few mm displaced from the optical axis. This reduces the magnet encumbrance, at expenses of a $\hat m - \hat e$ misalignment and consequent need of indirect gaze retrieval. Constructive efforts can be devoted to make the dipole parallel to the gaze direction (b) or perpendicular to it (c). This would simplify the data elaboration and --in the case (c), i.e. with a diametrical magnetization-- it would make the system highly responsive to torsional movements: an unprecedented feature of this methodology. As sketched, the (b) and (c) arrangements may present issues to guarantee a complete magnet embedding). \rosso{For simplicity, the angle \textit{kappa} is neglected in these figures: the optical and visual axes are represented as coincident. }
   }\label{fig:ideallens}
\end{figure*}

\rosso{A \textit{hardware} solution based on aligning the magnetic dipole to the visual axis of the eye would greatly simplify the task of retrieving the gaze from the measured $\hat m$. Unfortunately the visual axis is not known \textit{a priori}, and a patient based calibration would be anyhow needed. The visual axis is generally oriented at small angle from the optical axis (the so called angle \textit{kappa} amounts typically to about {5\degree} \cite{moshirfar_joph_13}), thus a compromise solution may consist in aligning  $\hat m$ to the optical axis and subsequently use the Listing's model described above for refinements. To this aim, the magnetic disc should be inserted out of the lens axis (to be non-obstructive for the view) and tilted with respect to the lens surface, an arrangement that comes with some geometrical issues, as sketched in Fig.\ref{fig:ideallens} b,c.}

In the scheme \ref{fig:ideallens} b, for instance, $\hat e_0$ and $\hat m_0$ come to a complete identity and no reciprocal rotation has to be measured or estimated. The scheme \ref{fig:ideallens} c, instead, opens new possibilities by allowing direct measurements of the torsional movement of the eye. These interesting perspectives are beyond the scope of the present work and will be subject of further investigation.

\subsection{Retrieving the eye gaze from the magnet position}
\label{subsec:r2gaze}

An alternative approach that could enable the estimation of the eye orientation with respect to the sensor array is based on the analysis of the magnet position. As said (Sec.\ref{subsec:bestfit}), the best fit procedure provides, beside $\vec B_{geo}$ and $\vec m$, also the position $\vec r$ of the dipole. 
As far as the eye is adequately described in terms of a fixed-centre sphere,  $\vec r$ is expected to move on a spherical surface. The radius of that sphere is the eye radius (12 mm, as a typical value), and its center is located at the eye center $\vec R_e$, which is (at least initially) an unknown position. 

\begin{figure*}[ht]
        \centering
        \includegraphics [angle=270, width=0.5\textwidth] {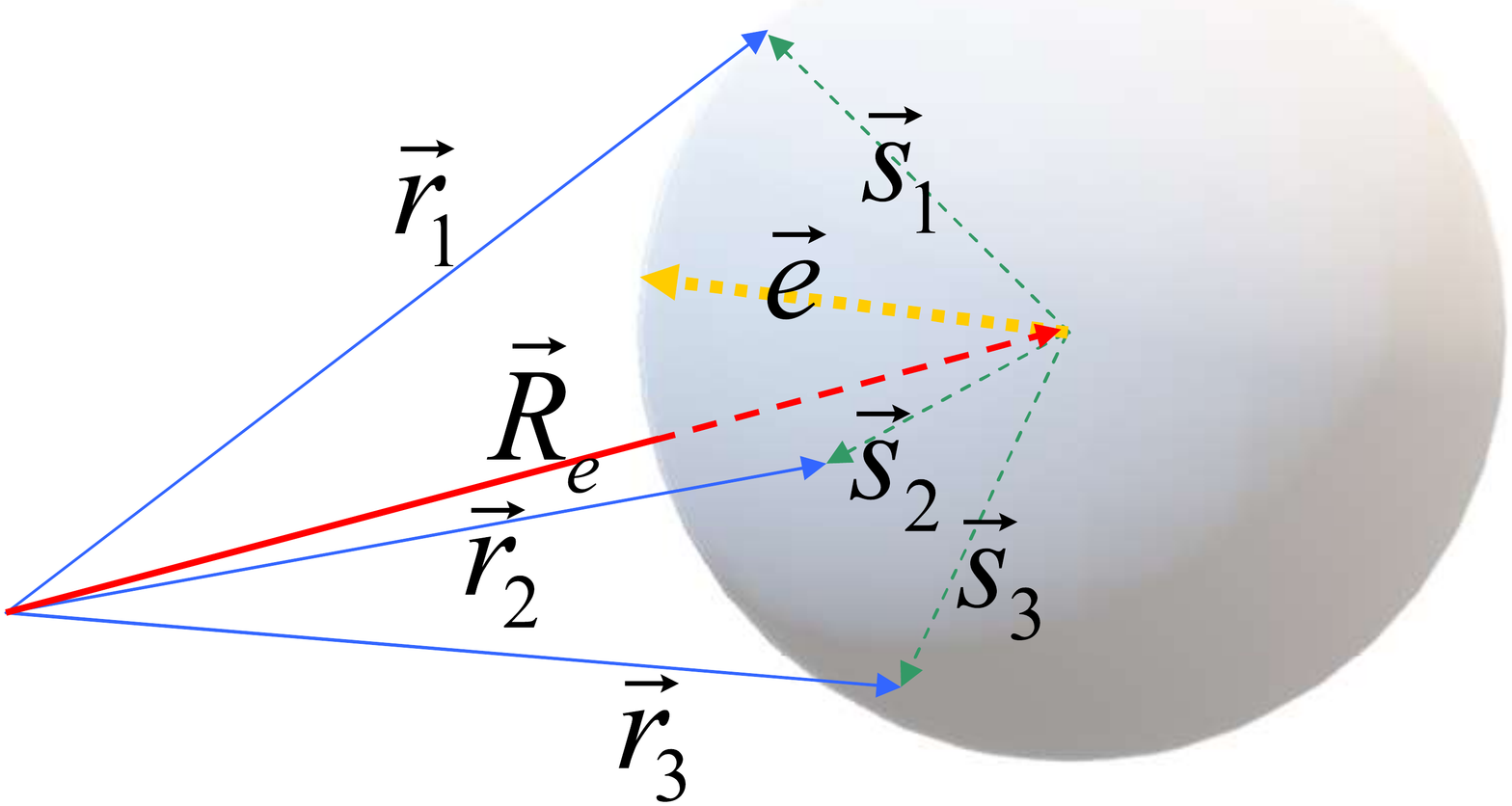}
  \caption{The retrieved positions {$\vec r_i$} are determined with respect to the sensor array frame. A large set  {$\vec r_i$} might enable the determination of the eye center $\vec R_e$, and hence the positions {$\vec s_i$} with respect to the eye center. The gaze orientation $\vec e$ moves rigidly with the position $\vec s$ of the magnet, and $\vec s$ could be used alternatively to (or in conjunction with) $\vec m$ to determine $\vec e$. This requires that $\vec R_e$ is determined with a high accuracy, which is not the case due to the small range of available {$\vec r_i$}.
  } \label{fig:rsRe}
\end{figure*}

In principle (see Fig.\ref{fig:rsRe}), one could first determine $\vec R_e$ fitting a large set of \{$\vec r_i$\} on a sphere and then use the quantities \{$\vec s_i = \vec r_i - \vec R_e$\}, similarly to what is done with $\vec m$.

More interestingly, provided that $\hat m$ e $\hat s$ are not parallel, it would be possible to use both to determine the eye orientation with no need for extra hypotheses. Indeed, $\hat m$,  $\hat s$ (e.g. together with $\hat m \times \hat s$) would constitute a basis in the 3D space, hence the matrix $R$ such that $\hat m_i = R\hat m_0$ and  $\hat s_i = R\hat s_0$ would be uniquely determined (even over-determined) from two subsequent estimates of $\hat m$ and $\hat s$.

Physiological constraints prevent $\vec r$ from ranging over a sufficiently  wide portion of the spherical surface and this results in a rather rough estimation of $\vec R_e$. As a consequence, these alternative approaches --at least with the precision achieved in current implementation-- do not help improve the quality of the gaze estimations. 

\rosso{An additional approach based on the analysis both $\vec r$ and $\vec m$  might follow the methodology presented  in Refs.\cite{guestrin_ieee_06, barsingerhorn_brm_18} for video-oculography, where the gaze is determined on the basis of the position of the pupil and of the eye glints, after opportune calibration. In the present case, apart from offsets to be determined by an additional calibration, $\vec r$ and $\vec m$ would provide analogous information as the pupil position and glints, respectively.
}

\section{\rosso{Instrumental uncertainties}}
\label{sec:uncertainties}

\begin{figure*}[ht]
\centering
\includegraphics[angle=0, width=0.48\textwidth]{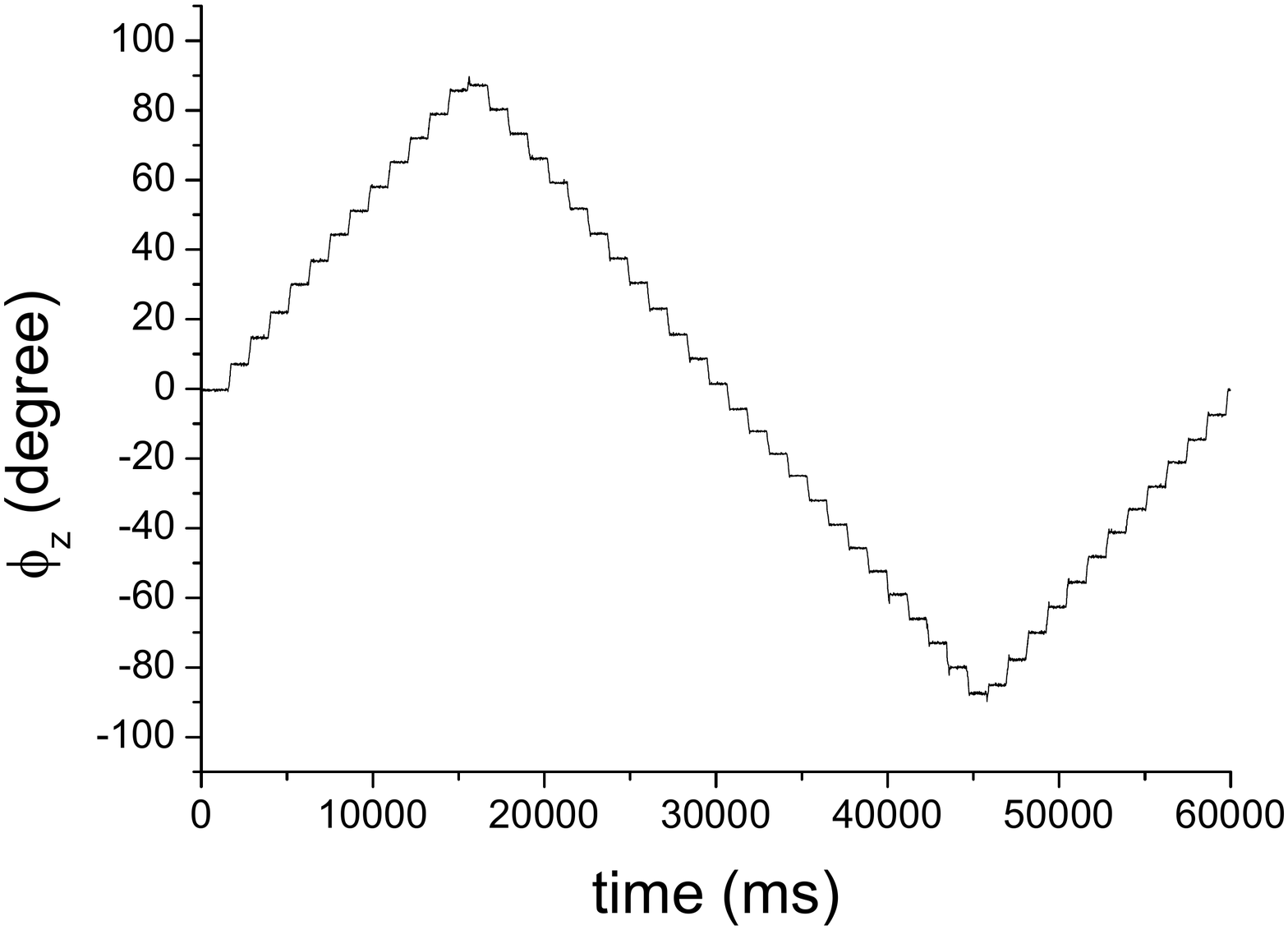}
\includegraphics[angle=0, width=0.48\textwidth]{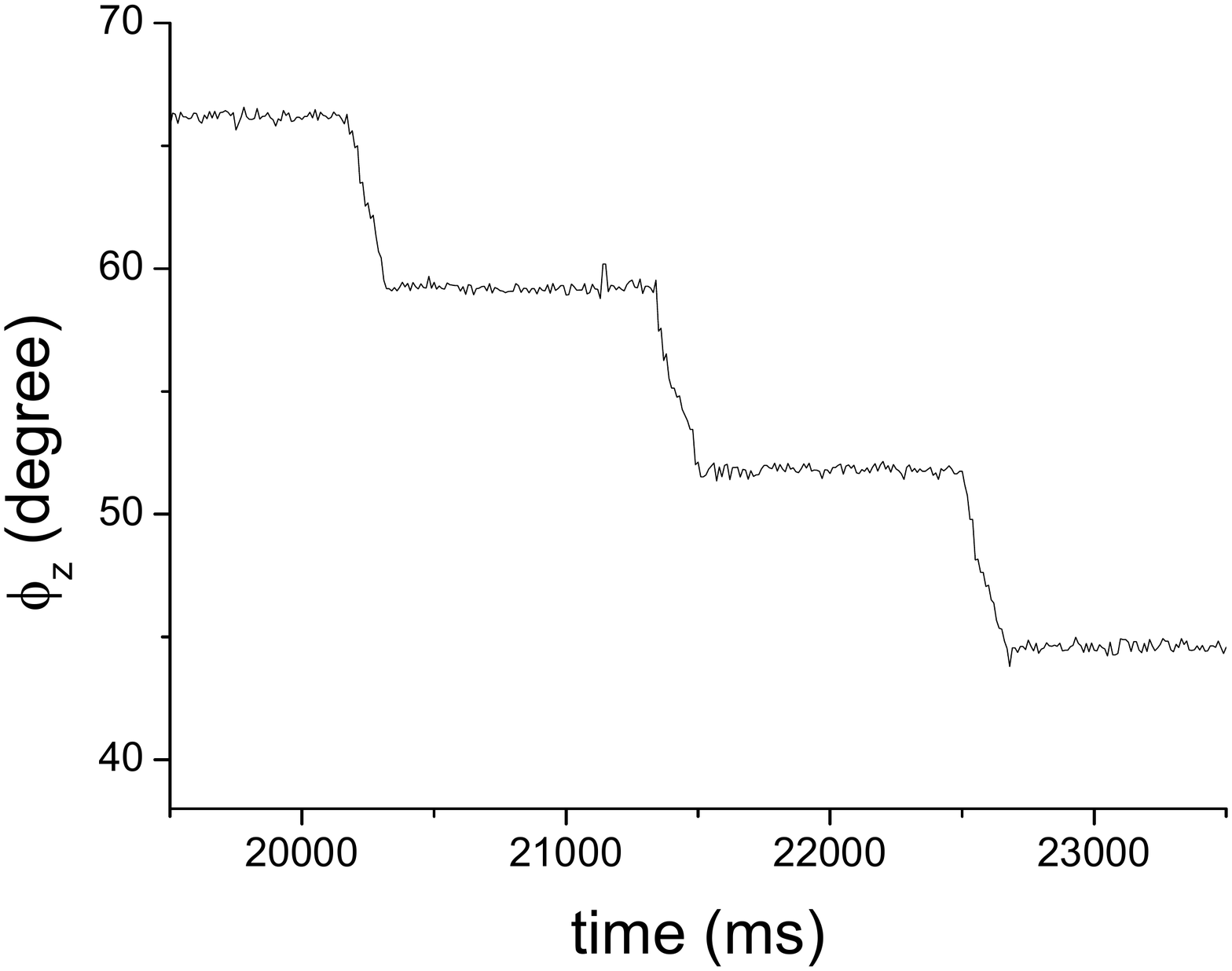}

  \caption{\rosso{The magnet rotates in steps around the $z$ axis on a circumference 8 mm in radius, each step corresponds to 360/51=7.06\degree and is followed by 100 trackings in static conditions. The non-monotonic behavior of the trace in the first plot is caused by the arcsin inversion, and can be numerically overcome. The zoomed plot at right provides a visual estimation of the precision, while the ramp (non)linearity accounts for the accuracy limits.}} \label{fig:stepangle}
  
\end{figure*}

\begin{figure*}[ht]
\centering
\includegraphics[angle=0, width=0.6\textwidth]{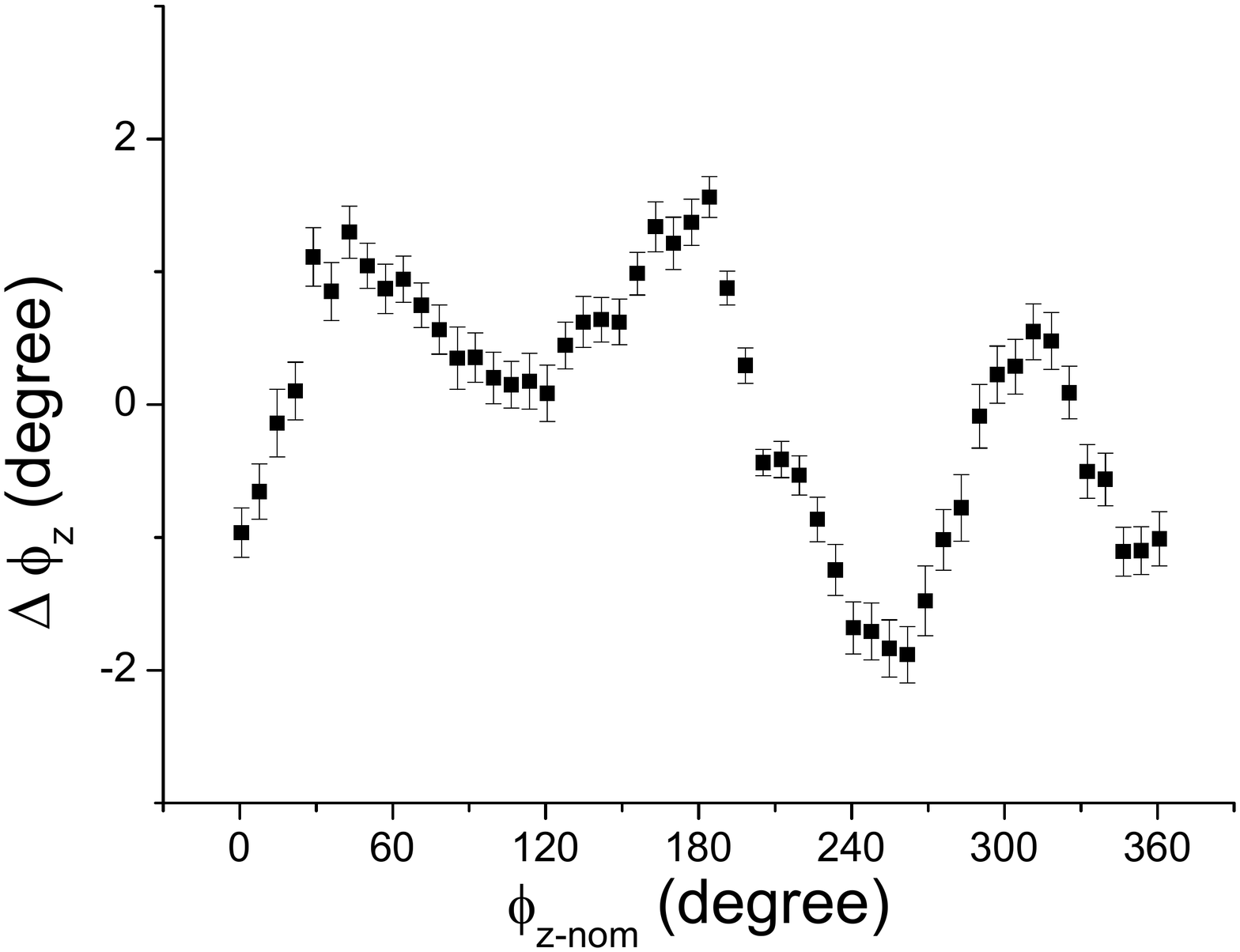}
  \caption{\rosso{The measurements plotted in Fig.\ref{fig:stepangle} are analyzed to quantify the angular uncertainties. Mean value and standard deviation of the estimated angle are evaluated after each step. The standard deviation is represented with the error bars, while  the mean value (corrected for non-monotonicity) is plotted  having subtracted of the nominal angle $\phi_{z-nom}$. The error bars provide an estimate of the instrumental precision, and the difference $\Delta \phi _z$ between estimated and nominal angles is a measure of the accuracy.} }\label{fig:uncertainties}
\end{figure*}

\rosso{
By operating with a single rotation axis, as described in \ref{subsec:oneaxis}, facilitates the quantification of 
the intrinsic instrumental precision and accuracy (that will affect the tracking, regardless of the methods used to infer $\hat e$ from $\vec m$), as set by the sensitivity of the detectors and by the operation conditions. To this end, we have performed several experiments aimed to provide estimates under typical operation conditions. The plots shown in Figs.\ref{fig:stepangle} and \ref{fig:uncertainties} are obtained from the elaboration of data recorded with a mechanical system based on a numerically controlled motor. The magnet is rotated in steps around the  $z$ axis along a circular orbit 8~mm in radius on the $z=10$mm plane (a region compatible with the eye positions), with the dipole radially oriented. A complete rotation (360\degree) is performed in 51 steps (7.06\degree each), and one hundred trackings are performed in static conditions after each rotation step.}

\rosso{
The data are then analyzed as described in Sec.\ref{subsec:param2eyegaze} and the rotation angle around the $z$ axis is inferred according to Eq.\ref{eq:arcsin}. The two plots in Fig.\ref{fig:stepangle} show the whole set of measurements and a zoomed portion of it, respectively.}

\rosso{
The plot in Fig.\ref{fig:uncertainties} reports the deviation from expected to measured angles, i.e. the difference between the average values of the angle retrieved at each step and the corresponding nominal angle (assigned by the stepped motor),  as a function of this latter. The error bars represent the standard deviation of the corresponding data sub-sets, and hence the precision. The non-zero value of the deviation exceeds the error bars, pointing out a limited accuracy due to systematic errors. These latter are likely to be ascribed to inhomogeneities of the ambient field and imperfect calibration of the sensors.}

\rosso{
The  above presented measurements were performed while maintaining the system steady with respect to the ambient field. This exclude the effect of covariance between different tracking parameters, which instead can be source of additional uncertainties.
Specifically, instrumental errors in the estimation of $\vec m$ can be evaluated when the whole system is rotated in the ambient field  and, complementarily, errors in the estimation of $\vec B_\mathrm{geo}$  can be evaluated while moving the magnet with respect to the steady sensors. Performing these experiments in our working conditions (operating in a normal room, with some care in avoiding ferromagnetic furniture in the proximity) we obtained typical RMS fluctuations of the order of {0.3\degree} within maximum ranges of about $\pm 1$\degree, for both the vectors, as detailed in Tab.\ref{tab:uncertainties} both in terms of RMS and maximal deviations.}

\begin{table}[ht]
    \centering
\rosso{   
\begin{tabular}{|c|c|c|}
       \hline
       quantity  &  st. dev.          & range                \\
       \hline
       $\phi_x$        &  0.23 \degree      & $\pm 0.98$\degree     \\        
       $\phi_y$        &  0.28 \degree      & $\pm 0.75$\degree     \\  
       $\theta_x$        &  0.33 \degree      & $\pm 0.50$\degree    \\  
       $\theta_y$        &  0.25 \degree      & $\pm 0.85$\degree    \\
       $\theta_z$        &  0.18 \degree      & $\pm 1.0$\degree    \\
       \hline
    \end{tabular}
    \caption {\rosso{The first two lines report the apparent rotation of $\vec m$ when the system is freely moved in the room and the magnet is in the $z = 10\,$mm plane, oriented along $z$, in a fixed position with respect to the sensor frame. Rotations around $x$ and $y$ are evaluated accordingly to eq.\ref{eq:arcsin} (the system is blind to rotations around $z$). The results are analyzed in terms of standard and maximal deviation. Conversely (last three lines), the apparent rotation of $\vec B_\mathrm{geo}$ is evaluated when the sensor array is fixed and the magnet (oriented along $z$) moves on a circle 10mm in radius on the $z = 10\,$mm plane. These estimates are just exemplifying numbers, as the results depend on several features of the working conditions (ambient field inhomogeneities,  accuracy of the sensor calibration,  free motion applied to the system) that assume typical but non-reproducible values.\label{tab:uncertainties}
    }}
}
\end{table}

\rosso{
It is important to take into account that other spurious phenomena may occur in the \textit{in-vivo} application and that they can relevantly affect the ultimate precision and accuracy. In particular, similarly to the case of SSC technology, slippage of the lens with respect to the eye constitute an error source. Also slippage of the sensor array with respect to the head must be carefully avoided. The latter is particularly relevant in head-impulse-test maneuvers for VOR-gain estimations. Future prototypes are being designed with a lightened sensor frame and a mechanically separated microcontroller PCB: a reduced inertia will help counteract such sensor-head slippage issue.
}

\rosso{
Furthermore, the rapid decay of the dipolar field with the distance, makes the magnet-sensor distance a crucial parameter to record data with a good signal to noise ratio. In \textit{in-vivo} application, physiological constraints hinder the task of maintaining the sensors at a short distance from the eye: an accurate design of the sensor frame is planned 
to this end, and it will greatly help improve the performance. 
}

\section{Conclusion}
\label{sec:conclusion}
An innovative eye-tracker is developed, based on non-inductive magnetometric measurements simultaneously performed at 100 Sa/s rate in a set of prearranged positions with the aid of an array of magnetoresistive sensors. This kind of instrumentation extends to the eye-tracking some advantageous features of similar magnetostatic trackers developed for other applications. 

Moderate invasivity, high speed, spatial and angular accuracy, robustness, simplicity and low cost emerge as attractive characteristics in comparison with other eye-tracking technologies.

Despite position and orientation of the magnet are fully characterized, the conversion of magnet tracking data to gaze information is not immediate and may require specific procedures. In this paper, we have examined the problem of retrieving the eye gaze from the magnet orientation when eye and dipole are not parallel.  

In particular, we have demonstrated how the combination of tracking parameters with physiological constraints of the eye motion enables an accurate reconstruction of the eye-gaze trajectories, while alternative simplified analyses can be implemented and --despite some distortion effects--  they enable the extraction of basic features of the gaze trajectory.

\section{Acknowledgments}
\label{sec:ack}
We are pleased to thank M.Mandalà, M.Carucci, A.Donniacuo and F.Viberti for useful discussions. Particular gratitude is expressed to M.Carucci who actively participated in the in-vivo measurements, wearing the lens and the tracker system. Those in-vivo measurements and preliminary ones were recorded with the help of A.Donniacuo, F.Viberti and L.Bellizzi. We also acknowledge the help of Y.Dancheva, who carefully read the manuscript and suggested important improvements.

\section*{Informed Consent}
{Informed consent was obtained from all subjects involved in the study. In particular, the system was worn by a scientist who actively participates in these studies and in this research.}

\bibliography{trck}

\end{document}